\documentclass{phep}

\journal{PAIML-ID}

\usepackage{amsmath}
\usepackage{amssymb}
\usepackage{graphicx}
\usepackage[hidelinks]{hyperref}

\newif\ifdraftversion
\draftversionfalse

\ifdraftversion
\usepackage{xcolor}
\usepackage{eso-pic}
\usepackage{tikz}

\AddToShipoutPictureFG{%
  \AtPageCenter{%
    \begin{tikzpicture}[remember picture, overlay]
      \foreach \x in {-10, 0, 10} {
        \foreach \y in {-12, -6, 0, 6, 12} {
          \node [rotate=45, shift={(\x cm, \y cm)}] {
            \fontfamily{phv}\selectfont
            \fontsize{45}{50}\selectfont
            \color{gray!60}
            \pgfsetfillopacity{0.25}
            \textbf{DRAFT}
          };
        }
      }
    \end{tikzpicture}%
  }%
}
\fi

\def\be{\begin{equation}}
\def\ee{\end{equation}}
\def\bea{\begin{eqnarray}}
\def\eea{\end{eqnarray}}

\renewcommand{\thefootnote}{\fnsymbol{footnote}}

\begin{document}

\title{Data-driven analysis of muon flux variations 
in muography time series at the Sos Enattos Mine}

\author{
M. Tramontini\auno{1}\footnotemark,
J.-C. Ianigro\auno{1},
J. Marteau\auno{1} 
}

\address{
\begin{flushleft}
$^1$Institut de Physique des 2 Infinis de Lyon, CNRS-IN2P3, UMR 5822, Université de Lyon, Université Claude Bernard Lyon 1, France\\

\end{flushleft}
}

\begin{abstract}

\begin{flushleft}
Muon flux variations in muography time series can provide information
on density-related changes in geological targets. A common approach is
to define regions of interest before constructing flux time series, but
this choice may dilute coherent signals or mask localized variations.
We present a data-driven analysis of two muon data acquisitions at the
Sos Enattos Mine in Sardinia, Italy. The method
aggregates lines of sight into overlapping spatial blocks, computes
standardized flux deviations, and applies principal component analysis
through singular value decomposition. The leading components reveal
spatially coherent structures and distinct temporal evolutions. In each
run, the first mode explains more than 20\% of the variance, while the
first eight modes explain approximately 80\%.
Regions selected from the spatial modes show anticorrelated and delayed
flux variations. These results show that the proposed
approach can identify candidate regions of coherent muon-flux
variability without imposing them a priori.

\par\vspace{8pt}


\par\vspace{4pt}

\noindent\textit{DOI:} -

\end{flushleft}
\end{abstract}

\maketitle

\begin{NoHyper} 
\renewcommand{\thefootnote}{\fnsymbol{footnote}}
\footnotetext{Corresponding author: m.tramontini@ip2i.in2p3.fr}
\end{NoHyper}

\section{Introduction}
Muography is a geophysical technique that uses cosmic-ray muons to
image the internal structure of large geological objects, such as
geological formations, archaeological sites
and volcanoes \citep{Bonechi2020}.
Muons are highly penetrating subatomic particles that can traverse significant
amounts of matter, allowing for the detection of density variations
within the target object. By measuring the flux of muons traversing
the object from different directions, it is possible to reconstruct a
muon image of the average density of its internal structure \citep{Tramontini2026}.
Furthermore, repeated flux measurements over time allow muography to
monitor temporal variations in the density of the target object, a
capability commonly referred to as muon monitoring \citep{Jourde2016}.

This monitoring capability is directly relevant to the Sos Enattos
mine, in Sardinia, Italy, a candidate site for the Einstein Telescope
(ET), a next-generation underground gravitational-wave detector.
Former mining sites such as Sos Enattos can be subject to gradual
changes in local hydrology and mass distribution, for instance
through groundwater infiltration or drainage within the gallery
network. Characterizing such mass redistribution contributes to the
broader geological and hydrological knowledge of the site.
Muography offers a means to directly probe these density variations 
from within the gallery, extending the range of techniques 
available for this purpose \citep{Tramontini2024,LeGonidec2019}.

One way of extracting temporal information from muography data is to
group individual lines of sight into a small number of regions of
interest, based on prior geological 
and/or geophysical knowledge, 
before computing a flux time series for each region. 
This a priori grouping choice can influence the resulting time
series and, in turn, the conclusions drawn from it. Relevant
coherent structures may be diluted by grouping together lines of
sight with unrelated behavior. Within a region, lines of sight with
a comparatively high muon rate can also dominate the aggregated
signal, masking the contribution of lower-rate lines of sight where
a relevant density variation may be located. A relevant region may
also be masked entirely if it is not among those selected.

In this work, we present results from two muon data
acquisitions at the Sos Enattos mine.
We introduce a principal-component-analysis
(PCA) based statistical method to identify coherent spatiotemporal
structures in the muon flux time series across many lines of sight,
without requiring an a priori choice of regions of interest. This
data-driven approach avoids the biases associated with a manual
grouping choice and lets the coherent structures emerge directly
from the data. We use this approach to isolate spatial regions 
exhibiting correlated behavior.

\section{Sos Enattos Mine}
The Sos Enattos mine is a former lead and zinc mine located
in the Nuoro province of Sardinia, Italy. The site
hosts an extensive network of underground galleries, developed
during the mine's operational history, which now provides access
for scientific instrumentation at various depths below the surface.
It also hosts the SarGrav laboratory, 
which provides underground facilities for geophysical 
and fundamental-physics experiments \citep{Saccorotti2023}.

The site is a candidate location for the Einstein Telescope (ET),
a next-generation underground gravitational-wave detector \citep{Punturo2010}.
Its low seismic background noise and the low population density of the surrounding area,
associated with reduced anthropogenic noise,
make Sos Enattos a suitable environment for such an instrument.
Dedicated site-characterization campaigns have been carried out to assess its geological
and environmental properties, 
including its seismic noise conditions \citep{Naticchioni2020,DiGiovanni2023}.
The pre-existing gallery network additionally offers convenient
underground access for deploying and operating geophysical
instrumentation, such as the muon detector used in this study
and a gaseous muon tracker developed by the HUN-REN Wigner Research Centre for Physics.

\section{Methodology}
\label{sec:methodology}

\begin{figure}[ht]
   \centering
   \includegraphics[width=\columnwidth]{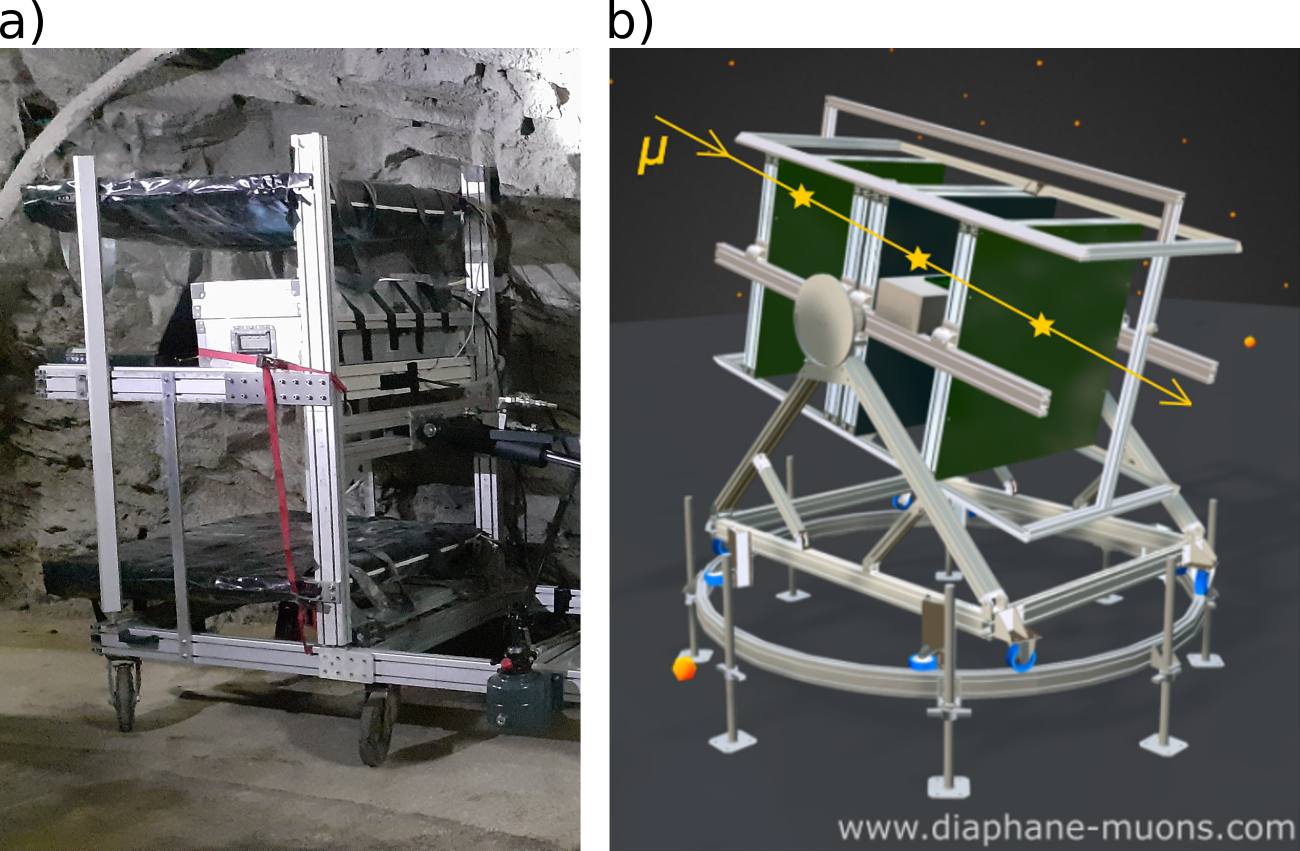}
   \caption{a) Muon detector installed in the Sos Enattos mine gallery.
   b) Schematic view of the muon detector.
   The muon trajectories (yellow straight line) are recovered from fired pixels (yellow stars).
   Modified from \citep{Tramontini2026}.}
   \label{fig:methodology:detector}
\end{figure}

The muon detector used in this study was developed by the
Institut de Physique des 2 Infinis de Lyon (IP2I) \cite[e.g.,][]{Marteau2012,Marteau2017}.
The instrument comprises three scintillator tracking planes, 
each formed by orthogonal arrays of 32 plastic scintillator bars,
providing a $32\times32$ spatial segmentation with a pixel size of $2.5\times2.5$ cm${}^2$.
The outermost planes are separated by 120 cm (Figure \ref{fig:methodology:detector}A).

Events are identified when signals are recorded in temporal coincidence across
the three tracking planes. The positions of the fired pixels are then
used to reconstruct the incident particle trajectory (Figure \ref{fig:methodology:detector}B).
We apply the RANSAC-based reconstruction and filtering procedure described in \cite{Bajou2023},
which determines the best-fitting straight-line trajectory while allowing spurious
or incompatible hits to be identified as outliers. 
Events for which a unique and sufficiently constrained trajectory cannot be established
are excluded from the analysis.
Additional information on the detector architecture and 
its optoelectronic readout is available in \cite{Marteau2014}.

\begin{figure}[ht]
   \centering
   \includegraphics[width=\columnwidth]{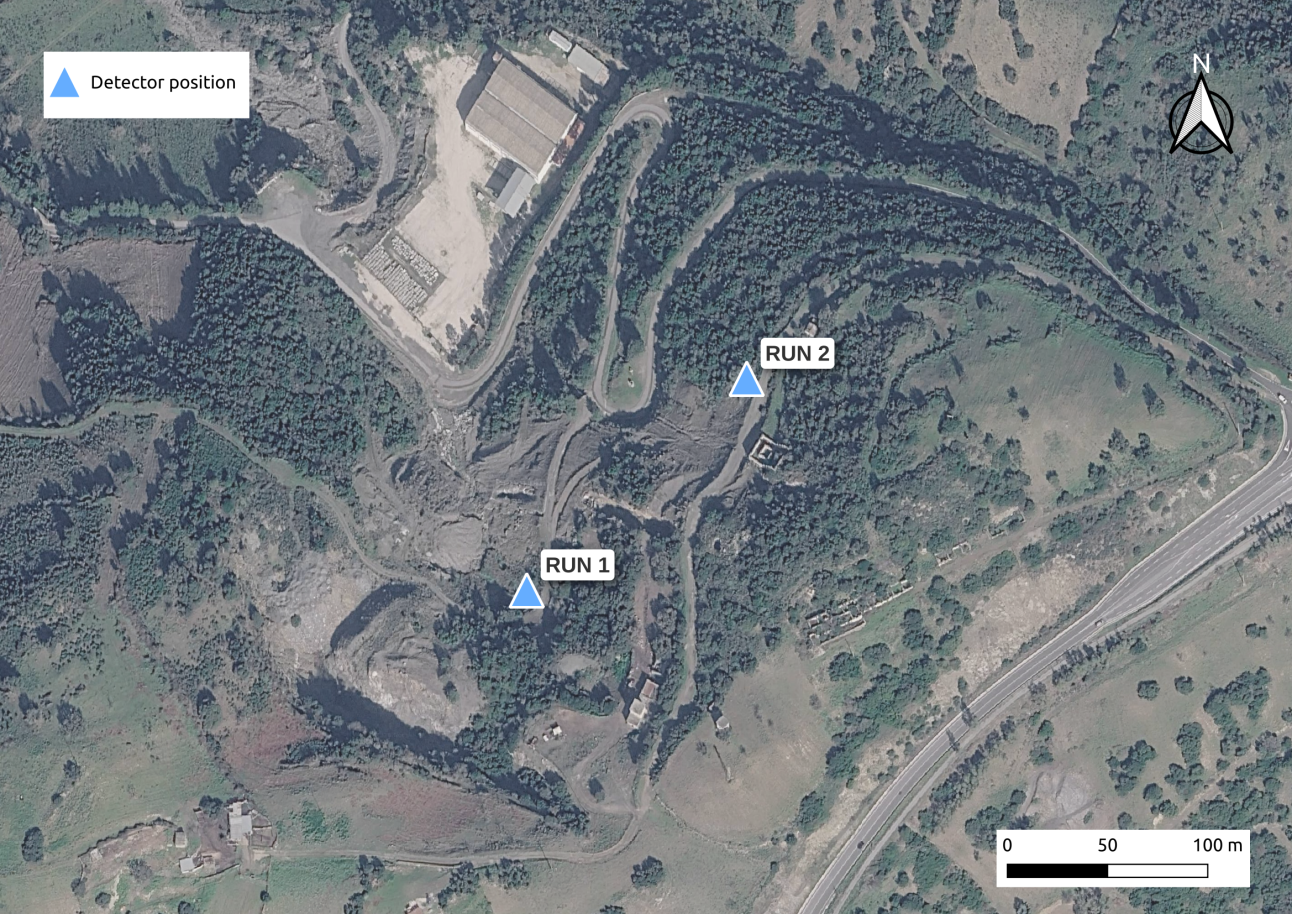}
   \caption{Satellite image showing the locations of the muon detector sites (RUN1 and RUN2) in the Sos Enattos mine gallery.}
   \label{fig:methodology:map}
\end{figure}

The detector was installed on two locations in the gallery (Figure \ref{fig:methodology:map}).
The selected sites, located at an altitude of  $\sim$220 m and $\sim$225 m,
offered a point to scan their rock overburden. These positions lie $\sim$145 m
and $\sim$170 m below the surface, respectively, and are separated by $\sim$150 m along the gallery.
For RUN1, the detector was oriented at $81.5^\circ$N and inclined at $0^\circ$.
For RUN2, the detector was oriented at $313^\circ$N and inclined at $0^\circ$.
Due to the settings used, the range of rock length traversed by the muons 
is 140-260 m and 155-260 m for RUN1 and RUN2, respectively.

Given the spatial extent of the structure investigated here,
the detector can be treated as a point-like instrument. 
Within this approximation, events sharing the same
relative direction between the front and rear matrices are assigned
to the same trajectory \cite{Lesparre2010}.
This yields \mbox{$(2N_x-1)(2N_y-1)=3969$} distinct lines of sight,
denoted $\hat{r}_{i,j}$. Here, 
\mbox{$(i,j)=(x,y)_\mathrm{front}-(x,y)_\mathrm{rear}$}
represents the difference between the pixel coordinates recorded
in the front and rear matrices, respectively. 
Each pair $(i,j)$ therefore defines a unique observation direction.

We define the integrated muon flux time series
$I_\mu(\hat{r}_{i,j},t)$ as the flux above the energy threshold imposed by
the crossed opacity. The direction-dependent acceptance
$\mathcal{T}(\hat{r}_{i,j})$ accounts for detector geometry and efficiency.
It is estimated from an open-sky calibration by comparing the observed rate
with the expected integrated open-sky flux,
$I^{\mathrm{open\mbox{-}sky}}(\hat{r}_{i,j})$.
We compute $I^{\mathrm{open\mbox{-}sky}}(\hat{r}_{i,j})$
using the parametrisation of Guan et al.~\citep{Guan2015}, corrected for
the detector altitude using Pumas \citep{Niess2022}, assuming a continuous
energy loss in the air.

We compute $I_\mu(\hat{r}_{i,j},t)$ as:
\begin{equation}
I_{\mu}(\hat{r}_{i,j},t)
=
\frac{N_{\mu}(\hat{r}_{i,j},t)}
{\mathcal{T}(\hat{r}_{i,j})\Delta T},
\end{equation}
where $N_{\mu}(\hat{r}_{i,j},t)$ is the number of detected muons along $\hat{r}_{i,j}$ and $\Delta T$ is the
acquisition duration.
We smooth the resulting time series using a 15-day Hamming 
moving-average window \citep{hamming1998digital}.
The window length was chosen to provide 
a balance between reducing short-term statistical fluctuations
and preserving temporal variations on the timescales
relevant to the analysis.

\subsection{Principal Component Analysis of the Integrated Flux Time Series}
\label{sec:pca}

To investigate the temporal evolution of $I_\mu(\hat{r}, t)$ and identify 
coherent patterns of variability, individual lines of sight $\hat{r} \equiv 
\hat{r}_{i,j}$ are aggregated into localized spatial blocks $b$.

Each block is indexed by its position
$b=(b_{\Delta x},b_{\Delta y})$ on the line-of-sight grid. It
contains an $s \times s$ contiguous set of adjacent directions
$\hat{r}$. We use an overlapping spatial configuration where blocks
are stepped by single line-of-sight increments, yielding a total of
$ N_b = (2N_x - s)(2N_y - s)$ spatial blocks.
For a given block, the integrated flux is
\begin{equation}
I_\mu(b,t)
=
\sum_{\hat{r} \in b} I_\mu(\hat{r},t).
\end{equation}
The corresponding uncertainty, $\sigma_I(b,t)$, is obtained by
combining the directional uncertainties in quadrature.
To assess the data quality of block $b$, direction-dependent
uncertainties are weighted by the total block flux and direction
coverage fraction,
\begin{equation}
Q_b = 
\frac{1}{f_b} 
\frac{\sqrt{\sum_{\hat{r} \in b} \sigma_{I,\hat{r}}^{2}}}
{\sum_{\hat{r} \in b} I_{\hat{r}}} \ ,
\end{equation}
where $I_{\hat{r}}$ and $\sigma_{I,\hat{r}}$ are the 
integrated muon flux and the corresponding
standard deviation for line of sight $\hat{r}$, and $f_b$ represents the 
fraction of valid directions within block $b$. 
Blocks yielding $Q_b > q_{\mathrm{th}}$ are excluded, where $q_{\mathrm{th}}$ 
is a predefined quality threshold. Near the field-of-view edges, $f_b$ 
naturally drops due to incomplete directional coverage.

For each accepted block, we compute the dimensionless standardized
flux deviation $Z(b,t)$ as
\begin{equation}
Z(b,t)
=
\frac{
I_{\mu}(b,t)-\left\langle I_{\mu}(b,t)\right\rangle
}{
\sigma_{I}(b,t)
},
\label{eq:sfd}
\end{equation}
where $\left\langle I_{\mu}(b,t)\right\rangle$ is the temporal mean of
the integrated flux for block $b$.
$\sigma_{I}(b,t)$ is the statistical uncertainty of the corresponding
flux estimate.

We compute the PCA modes via the singular value decomposition (SVD)
of $Z(b,t)$. To isolate coherent
spatiotemporal structures, we decompose the block-time field as
\begin{equation}
Z(b,t)
=
\sum_{k=1}^{K}\lambda_k u_k(b)v_k(t).
\end{equation}
Here, $u_k(b)$ is the $k$-th spatial mode and $v_k(t)$ is its temporal mode.
The singular value $\lambda_k$ measures the amplitude of the corresponding
dimensionless deviation structure. $K$ is the total number of modes.
We quantify the fraction of variance explained by the first $m$ modes
using the cumulative explained variance ratio (CEVR),
\begin{equation}
\mathrm{CEVR}(m)
=
\frac{\sum_{k=1}^{m}\lambda_k^2}
{\sum_{k=1}^{K}\lambda_k^2}.
\end{equation}

Because spatial blocks overlap and specific structural features may be 
expressed across multiple components, individual lines of sight $\hat{r}$ 
can receive overlapping mode assignments. 
To represent each spatial mode 
$u_k$ as a unique, full-resolution map over lines of sight $\hat{r}$, we 
resolve this ambiguity using a maximum-amplitude criterion.
Among all blocks $b$ containing direction $\hat{r}$, we retain the block
$b^\ast$ with the largest absolute spatial amplitude,
\begin{equation}
b^\ast = \operatorname*{arg\,max}_{b \, \ni \, \hat{r}} 
\left| u_k(b) \right| \ ,
\end{equation}
and assign its corresponding signed amplitude $u_k(b^\ast)$ to 
line of sight $\hat{r}$.

\section{Results and Discussion}
\label{sec:results}

\begin{figure}[ht]
   \centering
   \includegraphics[width=\columnwidth]{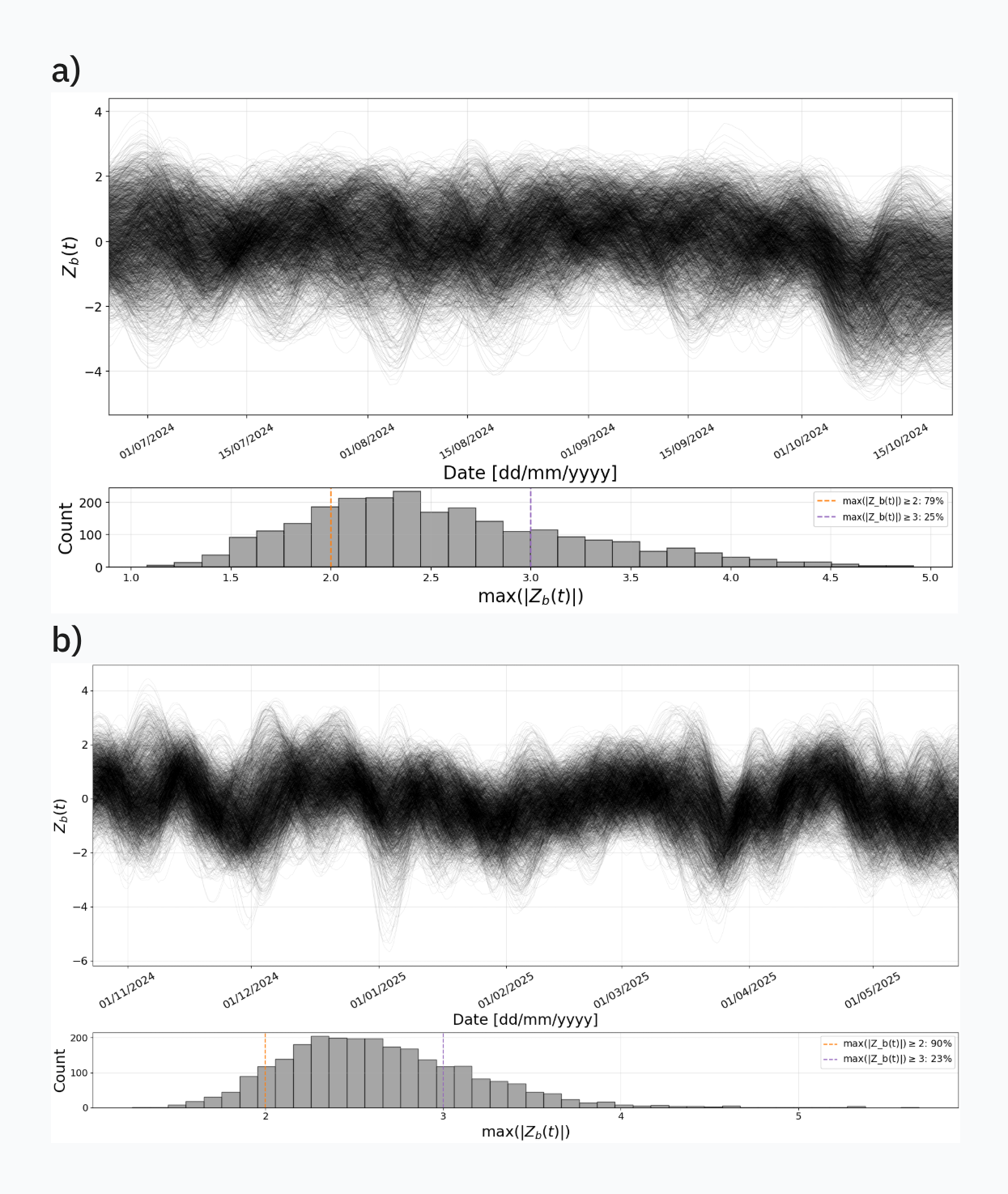}
   \caption{Standardized flux deviations $Z(b,t)$ for all valid spatial
   blocks in (a)~RUN1 and (b)~RUN2. In each panel, the upper plot shows
   the time series for all accepted blocks, with one curve per spatial
   block. The lower plot shows the distribution of the maximum absolute
   deviation reached by each block, $\max(|Z(b,t)|)$. Dashed vertical
   lines mark the thresholds at 2 and 3. The legend gives the fraction
   of blocks whose maximum absolute deviation exceeds each threshold.}
   \label{fig:res:signals}
\end{figure}

The block dimension and quality threshold
were selected to provide a balance
between spatial resolution,
statistical stability, and data quality.
We choose the block dimension to be $s = 10$, yielding a total of
$N_b = 2916$ spatial blocks for each RUN. We define the quality threshold
for excluding low-quality blocks as $q_{\mathrm{th}} = 0.05$. 
This resulted in 2436 and 2557 valid spatial blocks for RUN1 and
RUN2, respectively. The block-level quantities $Z(b,t)$ are shown in
Figure~\ref{fig:res:signals}. The lower histograms summarize the
largest absolute value reached by each block. In RUN1, 79\% and
25\% of the blocks reach $\max(|Z(b,t)|) \geq 2$ and $3$,
respectively. In RUN2, the corresponding fractions are 90\% and
23\%, respectively.

\begin{figure}[ht]
   \centering
   \includegraphics[width=\columnwidth]{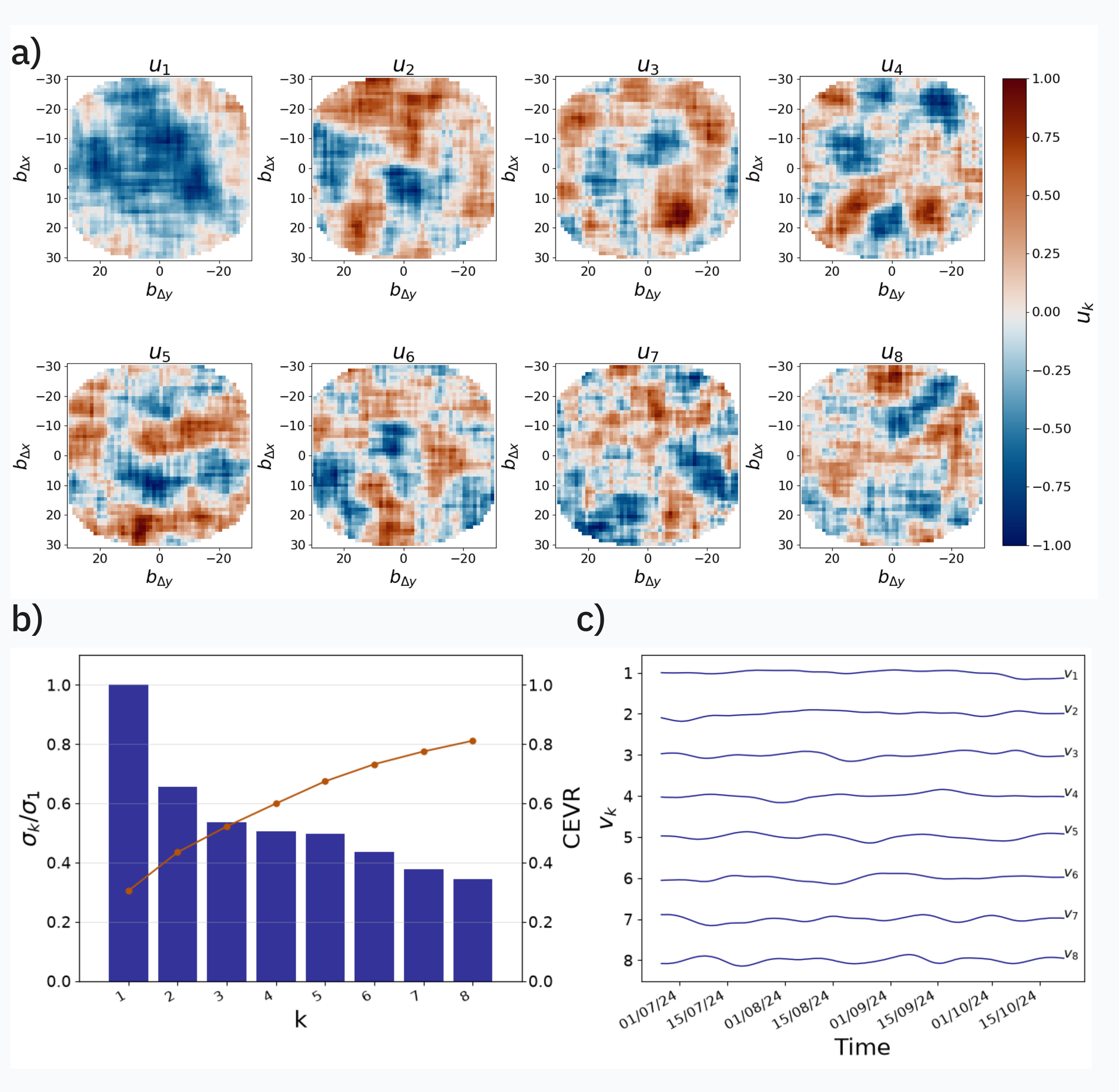}
   \caption{SVD of the standardized flux deviations for RUN1.
   (a) Spatial maps of
   the first eight modes $u_k(\hat{r})$ over the block grid, ordered
   by decreasing singular value. (b) Normalized singular values
   $\lambda_k/\lambda_1$ (bars, left axis) and cumulative explained
   variance ratio (line, right axis) as a function of mode order
   $k$. (c) Temporal modes $v_k(t)$, offset vertically for clarity.}
   \label{fig:res:svd_run1}
\end{figure}

\begin{figure}[ht]
   \centering
   \includegraphics[width=\columnwidth]{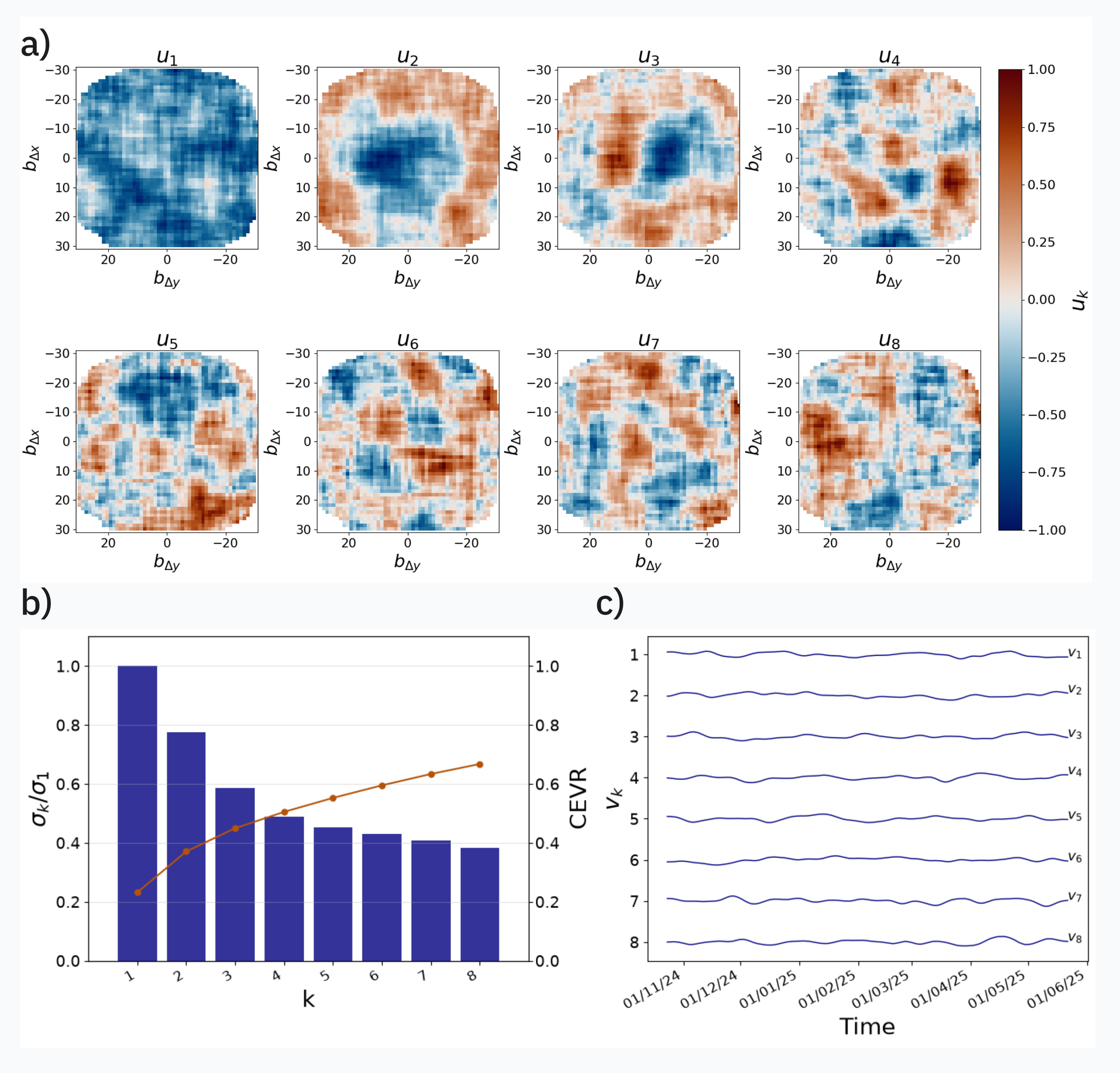}
   \caption{SVD of the standardized flux deviations for RUN2.
   (a) Spatial maps of
   the first eight modes $u_k(\hat{r})$ over the block grid, ordered
   by decreasing singular value. (b) Normalized singular values
   $\lambda_k/\lambda_1$ (bars, left axis) and cumulative explained
   variance ratio (line, right axis) as a function of mode order
   $k$. (c) Temporal modes $v_k(t)$, offset vertically for clarity.}
   \label{fig:res:svd_run2}
\end{figure}

The SVD results are presented in Figures~\ref{fig:res:svd_run1}
and~\ref{fig:res:svd_run2}.
Panel~(a) shows the first eight spatial
modes, $u_k(b)$, for $k=1,\ldots,8$. 
Perceptually uniform colour maps
are used to visualize the spatial modes 
and avoid visual distortion \citep{Crameri2020}.
Panel~(b) shows the
normalized singular values $\lambda_k/\lambda_1$ together with the
CEVR. Panel~(c) shows the corresponding temporal modes $v_k(t)$.
From the CEVR curves, the first mode accounts for about 30\% of the
variance in RUN1, while the first eight modes account for about
80\%. In RUN2, these values are about 23\% and 67\%, respectively.
In both RUNs, the singular-value spectrum decreases gradually. This
indicates that the variability is distributed over several coherent
modes rather than concentrated in a single dominant component.
In both RUNs, the first spatial mode is dominated by one sign over most
of the accepted field of view. This suggests that the leading component
mainly captures a common-mode variation shared by many spatial blocks.
Higher-order modes contain clearer positive and negative domains, which
separate regions with opposite contributions to the same temporal mode.
The temporal modes in panel~(c) show that these spatial patterns are
associated with distinct time evolutions during the acquisition period.

\subsection{Region Selection from Spatial Modes}

Beyond mapping individual modes $u_k$, we define spatial regions of interest
jointly from two chosen modes, $u_i$ and $u_j$, by taking into account
the corresponding singular values $\lambda_i$ and $\lambda_j$.
Each spatial 
block $b$ is represented as a point in the
$(\lambda_i u_i(b), \lambda_j u_j(b))$ projection
plane, and a region is obtained by selecting on this plane,
subject to minimum-magnitude 
thresholds $\tau_i$ and $\tau_j$ that exclude blocks with weak projections.

Because a given block $b$ may qualify for regions defined from different 
mode pairs, we resolve competing assignments using a 
maximum-amplitude criterion. Among the candidate modes
$k \in \{i, j\}$, block $b$ is assigned to the mode that maximizes
the singular-value-weighted amplitude,
\begin{equation}
k^\ast = \operatorname*{arg\,max}_{k \, \in \, \{i,j\}}
\left| \lambda_k \, u_k(b) \right| \ .
\end{equation}
Unlike the single-mode mapping used to render individual components 
in Section~\ref{sec:methodology} (where $k$ is fixed and $\lambda_k$
cancels from the comparison), here the singular values differ between 
the competing modes and therefore remain relevant. A block with a
comparatively large $u_i(b)$ may still be governed by mode $j$ if
$\lambda_j \gg \lambda_i$.

\begin{figure}[ht]
   \centering
   \includegraphics[width=\columnwidth]{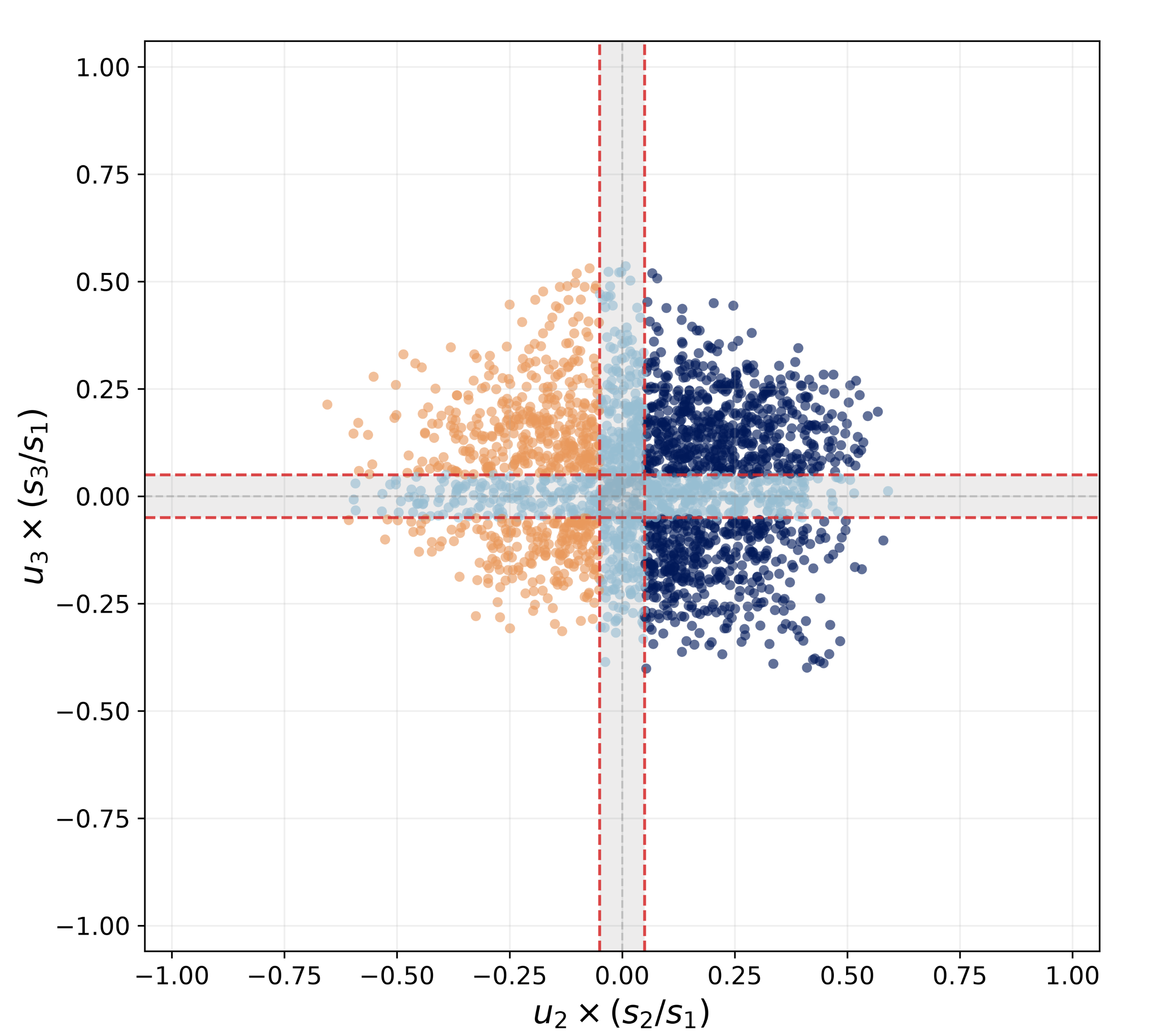}
   \caption{Region selection for RUN1 based on the spatial modes
   $u_2$ and $u_3$. Left panel shows spatial blocks in the
   ($\lambda_2 u_2(b)$, $\lambda_3 u_3(b)$) projection plane. Blocks
   assigned to Region~1 are shown in dark blue, blocks assigned to
   Region~2 are shown in orange, and blocks excluded by the
   minimum-magnitude thresholds $\tau_2$ and $\tau_3$ (dashed lines)
   are shown in light blue. Right panels show spatial maps of the
   weighted amplitudes $\lambda_2 u_2(b)$ (top) and
   $\lambda_3 u_3(b)$ (bottom) over the block grid, using the same
   region color coding.}
   \label{fig:dis:selection_run1}
\end{figure}

For RUN1, we apply the region selection procedure described above to
identify spatial regions of interest based on the spatial modes $u_2$ and $u_3$.
To exclude blocks with weak projections, 
we set a minimum-magnitude threshold of $\tau_2=\tau_3=0.05$.
We define Region 1 as the set of blocks 
for which $u_2(b)>0$ and $| \lambda_2 u_2(b) | > \tau_2$ and
$| \lambda_3 u_3(b) | > \tau_3$.
We define Region 2 as the set of blocks 
for which $u_2(b)<0$ and $| \lambda_2 u_2(b) | > \tau_2$ and
$| \lambda_3 u_3(b) | > \tau_3$.
The selection of blocks for Regions 1 and 2 is shown in
Figure~\ref{fig:dis:selection_run1}.
We map the selected regions back onto the lines of sight,
and compute the corresponding $I_\mu(t)$ for each selected region.
We apply Equation~\ref{eq:sfd} to each selected region.

For each run $r$ and region $i$,
$\langle I_{\mu,i}^{(r)}\rangle$ and $\sigma_{I,i}^{(r)}$ denote
the temporal mean and standard deviation of the integrated flux, respectively.
The corresponding standardized flux deviation is denoted by $Z_i^{(r)}(t)$.
As a standardized deviation, $Z_i^{(r)}(t)$ can be interpreted as a z-score, 
expressing the deviation of the regional flux 
from its temporal mean in units of its standard deviation.
The reference levels $|Z_i^{(r)}|=1$, $2$, and $3$ are shown on
the regional deviation plots, corresponding to one, two,
and three standard deviations from the mean, respectively.

\begin{figure}[ht]
   \centering
   \includegraphics[width=\columnwidth]{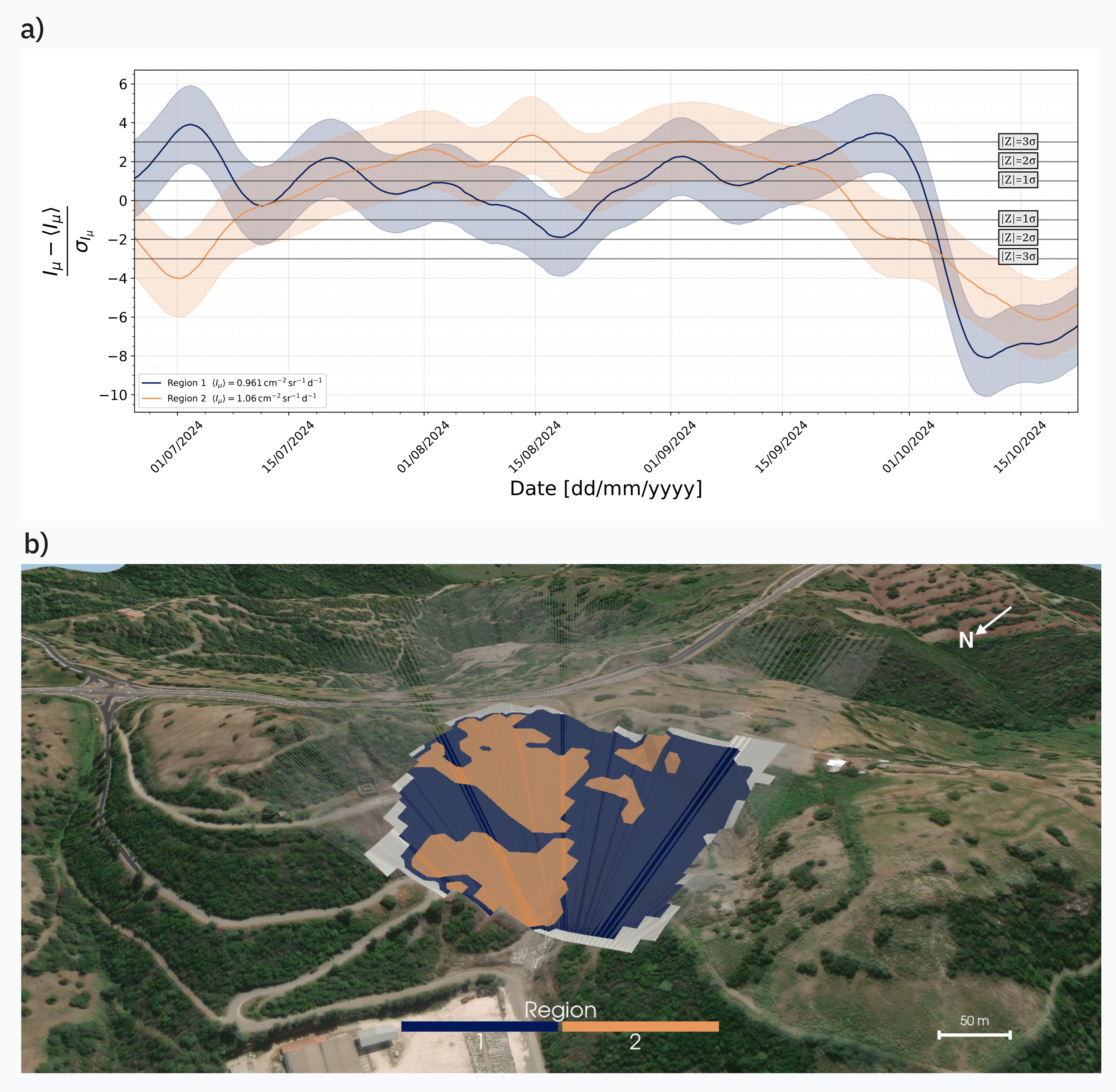}
   \caption{Results for the regions selected in RUN1. (a) Regional
   standardized flux deviations $Z_i^{(1)}(t)$ for Region~1
   (dark blue) and Region~2 (orange), with shaded bands showing the
   associated statistical uncertainty. Horizontal dashed lines mark
   the $|Z_i^{(1)}|=1$, $2$ and $3$ reference levels. The mean integrated
   flux $\langle I_\mu\rangle$ of each region is given in the
   legend. (b) Spatial extent of Region~1 (dark blue) and Region~2
   (orange), overlaid on a 3D view of the gallery.}
   \label{fig:dis:selection_run1_results}
\end{figure}

Figure~\ref{fig:dis:selection_run1_results}a shows $Z_1^{(1)}(t)$
and $Z_2^{(1)}(t)$. The spatial extent of each region is shown in
Figure~\ref{fig:dis:selection_run1_results}b.
$\langle I_{\mu,1}^{(1)}\rangle =
0.961~\text{cm}^{-2}~\text{sr}^{-1}~\text{d}^{-1}$ and
$\langle I_{\mu,2}^{(1)}\rangle =
1.06~\text{cm}^{-2}~\text{sr}^{-1}~\text{d}^{-1}$,
resulting in an average flux about 10\% higher in Region~2 than
Region~1. During the first half of the acquisition period,
$Z_1^{(1)}$ and $Z_2^{(1)}$ are anticorrelated. Between
mid-July and mid-August 2024, $Z_1^{(1)}$ decreases from about $+4$
to $-2$, while $Z_2^{(1)}$ increases from about $-4$ to $+3$.
Similar anticorrelated variations occur in early and late September
2024. From early October to the end of RUN1, both regions decrease
and reach negative extrema near $Z_1^{(1)}=-8$ and
$Z_2^{(1)}=-6$.

The anticorrelation episodes between the two regions
suggests that the observed variations 
are not purely common-mode fluctuations 
affecting the entire detector field of view.
Instead, the opposite evolution of the two regional fluxes
may indicate a spatially localized process
affecting the corresponding lines of sight.
A possible interpretation is a redistribution of mass 
between the two regions,
which would produce opposite variations in the measured integrated fluxes.
The anticorrelated episodes observed in RUN1 therefore
provide a first indication of localized changes
in the effective mass distribution within the monitored volume.

\begin{figure}[ht]
   \centering
   \includegraphics[width=\columnwidth]{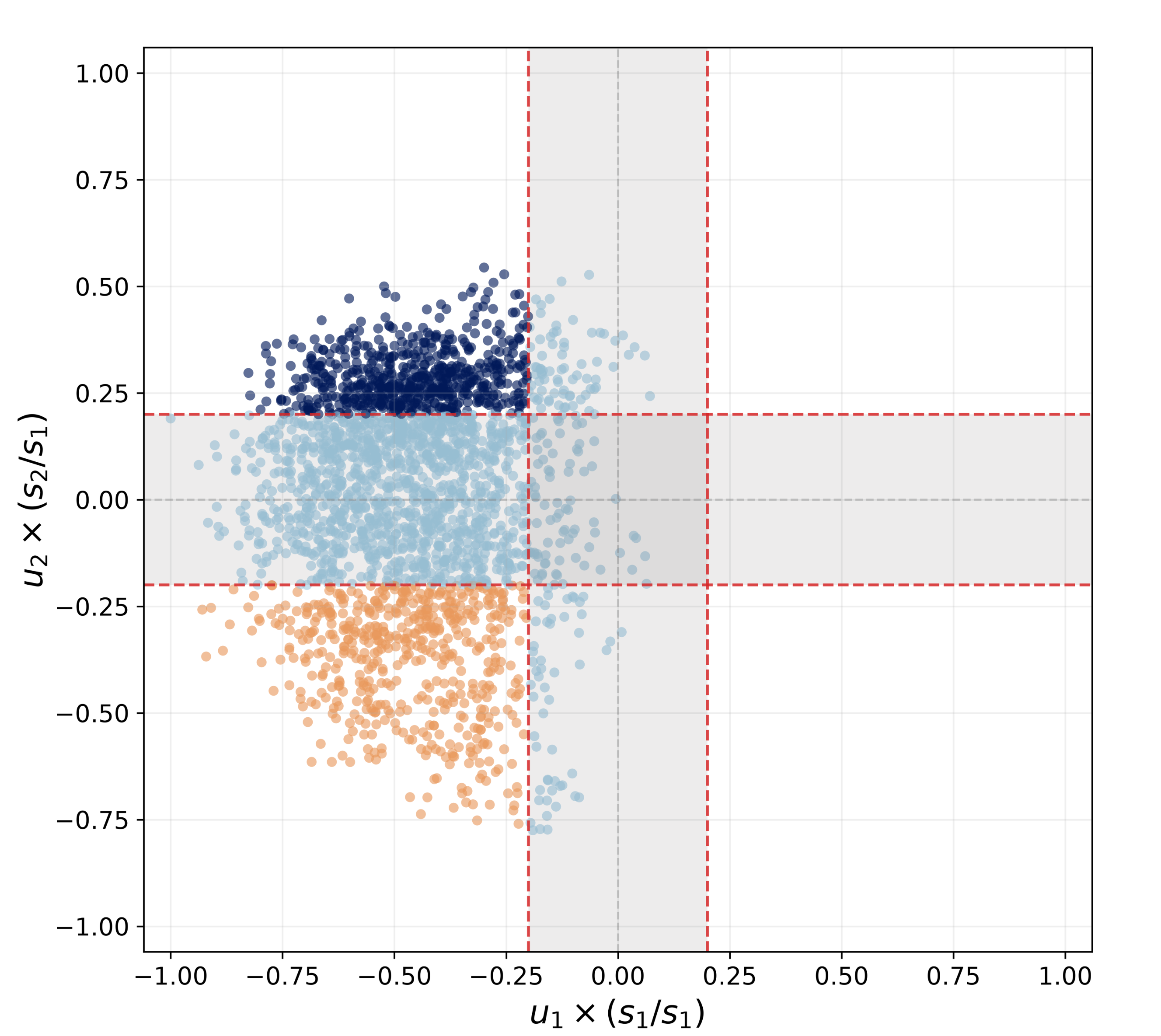}
   \caption{Region selection for RUN2 based on the spatial modes
   $u_1$ and $u_2$. Left panel shows spatial blocks in the
   ($\lambda_1 u_1(b)$, $\lambda_2 u_2(b)$) projection plane. Blocks
   assigned to Region~1 are shown in dark blue, blocks assigned to
   Region~2 are shown in orange, and blocks excluded by the
   minimum-magnitude thresholds $\tau_1$ and $\tau_2$ (dashed lines)
   are shown in light blue. Right panels show spatial maps of the
   weighted amplitudes $\lambda_1 u_1(b)$ (top) and
   $\lambda_2 u_2(b)$ (bottom) over the block grid, using the same
   region color coding.}
   \label{fig:dis:selection_run2}
\end{figure}

For RUN2, we apply the region selection procedure described above to
identify spatial regions of interest based on the spatial modes $u_1$ and $u_2$.
To exclude blocks with weak projections, 
we set a minimum-magnitude threshold of $\tau_1=\tau_2=0.2$.
We define Region 1 as the set of blocks 
for which $u_2(b)>0$ and $| \lambda_1 u_1(b) | > \tau_1$ and
$| \lambda_2 u_2(b) | > \tau_2$.
We define Region 2 as the set of blocks 
for which $u_2(b)<0$ and $| \lambda_1 u_1(b) | > \tau_1$ and
$| \lambda_2 u_2(b) | > \tau_2$.
The selection of blocks for Regions 1 and 2 is shown in
Figure~\ref{fig:dis:selection_run2}.

\begin{figure}[ht]
   \centering
   \includegraphics[width=\columnwidth]{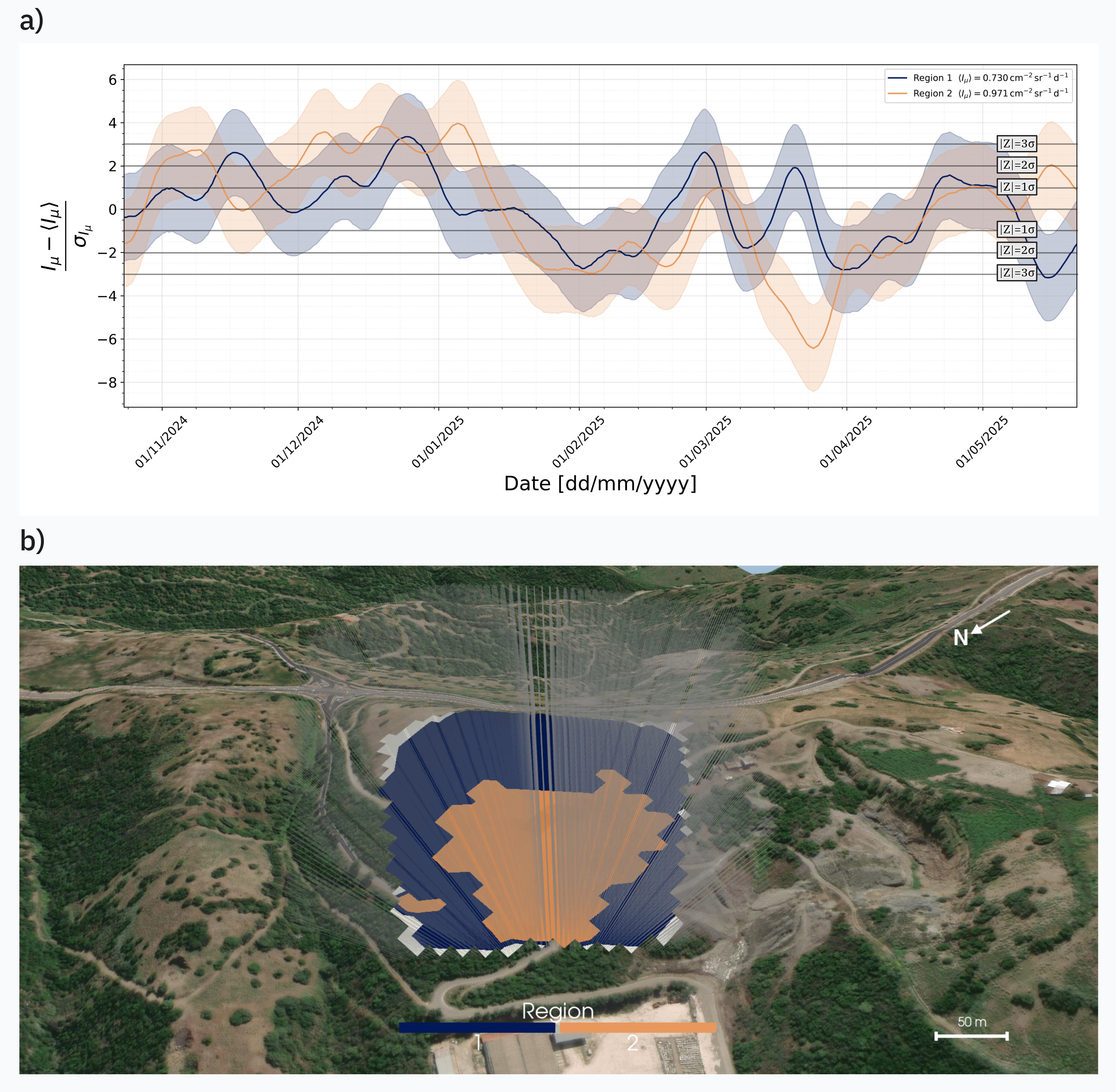}
   \caption{Results for the regions selected in RUN2. (a) Regional
   standardized flux deviations $Z_i^{(2)}(t)$ for Region~1
   (dark blue) and Region~2 (orange), with shaded bands showing the
   associated statistical uncertainty. Horizontal dashed lines mark
   the $|Z_i^{(2)}|=1$, $2$ and $3$ reference levels. The mean integrated
   flux $\langle I_\mu\rangle$ of each region is given in the
   legend. (b) Spatial extent of Region~1 (dark blue) and Region~2
   (orange), overlaid on a 3D view of the gallery.}
   \label{fig:dis:selection_run2_results}
\end{figure}

We map the selected regions back onto the lines of sight,
and compute the corresponding $I_\mu$ for each selected region.
Figure~\ref{fig:dis:selection_run2_results}a shows $Z_1^{(2)}(t)$
and $Z_2^{(2)}(t)$. The spatial extent of each region is shown in
Figure~\ref{fig:dis:selection_run2_results}b.
$\langle I_{\mu,1}^{(2)}\rangle =
0.730~\text{cm}^{-2}~\text{sr}^{-1}~\text{d}^{-1}$ and
$\langle I_{\mu,2}^{(2)}\rangle =
0.971~\text{cm}^{-2}~\text{sr}^{-1}~\text{d}^{-1}$, resulting
in an average flux about 30\% higher in Region~2 than Region~1.

The temporal variations $Z_1^{(2)}(t)$ and $Z_2^{(2)}(t)$
exhibit a relative delay during two intervals.
From early November 2024 to near late
December 2024, variations in Region~2 precede similar variations in
Region~1 by approximately 10 days,
whereas between mid-February and early March
2025, the delayed pattern is reversed, with Region~1 leading Region~2.
Additional localized anticorrelations occur in late March 2025,
where $Z_1^{(2)}$ is around $+2$ while $Z_2^{(2)}$ falls below
$-6$, and in mid-May 2025, where $Z_1^{(2)}$ is about $-3$ while
$Z_2^{(2)}$ is about $+2$.

The observed delayed responses suggest that the temporal variations
are not simultaneous across the monitored volume. The reversal in
the direction of the delay between the two periods may indicate
that the temporal response of different regions is not uniform.
Together with the localized anticorrelations observed later in the
acquisition period, these results suggest that the temporal
variations detected by the muon flux are spatially heterogeneous
and may reflect processes occurring at different locations or with
different characteristic timescales within the monitored volume.

\section{Conclusions}
\label{sec:conclusions}
We presented a PCA-based workflow for analyzing muography time series
without imposing regions of interest a priori. The method first
standardizes the integrated flux in overlapping spatial blocks, then
uses SVD to identify coherent spatial and temporal modes. This provides
a way to define regions from the data while retaining the sign and
relative amplitude of the dominant components.

The application to two muography datasets at the Sos Enattos Mine shows
that the variability is distributed over several components. The CEVR
indicates that the first mode explains about 30\% of the variance in
RUN1 and about 23\% in RUN2. The first eight modes explain about
80\% and 67\%, respectively. 
Regions selected from these modes
exhibit distinct temporal relationships. In RUN1, the two regions show
episodes of both correlated and anticorrelated variations, with the
latter suggesting spatially heterogeneous flux variations. In RUN2,
the dominant relationship is characterized by delayed responses:
variations in one region precede similar variations in the other by
approximately 10 days during two periods, with the direction of the
delay reversed between them. Additional episodes of localized
anticorrelation are also observed.

These results demonstrate that the proposed method can identify
spatial regions associated with distinct patterns of muon-flux
variability, including correlated, anticorrelated, and temporally
delayed responses. The approach therefore provides a data-driven
framework for investigating spatially heterogeneous temporal
variations in muography time series.

\section*{Conflict of Interest}
The authors declare that there are no conflict of interest
regarding the publication of this paper.

\section*{Acknowledgments}
We thank Irene Fiori and Nicolas Arnaud (VIRGO collaboration) 
for discussions on the noise environment around the VIRGO detector,
the staff at the Sos Enattos Mine for their logistical support on site,
and Andrea Contu and Domenico D'Urso for coordinating the experiment.

\bibliographystyle{phep}
\bibliography{bib-file}

@article{Crameri2020,
  title = {The misuse of colour in science communication},
  volume = {11},
  ISSN = {2041-1723},
  url = {http://dx.doi.org/10.1038/s41467-020-19160-7},
  DOI = {10.1038/s41467-020-19160-7},
  number = {1},
  journal = {Nature Communications},
  publisher = {Springer Science and Business Media LLC},
  author = {Crameri,  Fabio and Shephard,  Grace E. and Heron,  Philip J.},
  year = {2020},
  month = Oct 
}

@article{Niess2022,
  title = {The {PUMAS} library},
  volume = {279},
  ISSN = {0010-4655},
  url = {http://dx.doi.org/10.1016/j.cpc.2022.108438},
  DOI = {10.1016/j.cpc.2022.108438},
  journal = {Computer Physics Communications},
  publisher = {Elsevier BV},
  author = {Niess,  Valentin},
  year = {2022},
  month = Oct,
  pages = {108438}
}

@misc{Guan2015,
  doi = {10.48550/ARXIV.1509.06176},
  url = {https://arxiv.org/abs/1509.06176},
  author = {Guan,  Mengyun and Chu,  Ming-Chung and Cao,  Jun and Luk,  Kam-Biu and Yang,  Changgen},
  title = {A parametrization of the cosmic-ray muon flux at sea-level},
  publisher = {arXiv},
  year = {2015},
  copyright = {arXiv.org perpetual,  non-exclusive license}
}

@article{Jourde2016,
  title = {Monitoring temporal opacity fluctuations of large structures with muon radiography: a calibration experiment using a water tower},
  volume = {6},
  ISSN = {2045-2322},
  url = {http://dx.doi.org/10.1038/srep23054},
  DOI = {10.1038/srep23054},
  number = {1},
  journal = {Scientific Reports},
  publisher = {Springer Science and Business Media LLC},
  author = {Jourde,  Kevin and Gibert,  Dominique and Marteau,  Jacques and de Bremond d’Ars,  Jean and Gardien,  Serge and Girerd,  Claude and Ianigro,  Jean-Christophe},
  year = {2016},
  month = Mar 
}

@article{DiGiovanni2023,
  title = {Temporal variations of the ambient seismic field at the
    {Sardinia} candidate site of the {Einstein Telescope}},
  volume = {234},
  ISSN = {1365-246X},
  url = {http://dx.doi.org/10.1093/gji/ggad178},
  DOI = {10.1093/gji/ggad178},
  number = {3},
  journal = {Geophysical Journal International},
  publisher = {Oxford University Press (OUP)},
  author = {Di Giovanni,  M and Koley,  S and Ensing,  J X and Andric,  T and Harms,  J and D’Urso,  D and Naticchioni,  L and De Rosa,  R and Giunchi,  C and Allocca,  A and Cadoni,  M and Calloni,  E and Cardini,  A and Carpinelli,  M and Contu,  A and Errico,  L and Mangano,  V and Olivieri,  M and Punturo,  M and Rapagnani,  P and Ricci,  F and Rozza,  D and Saccorotti,  G and Trozzo,  L and Dell’aquila,  D and Pesenti,  L and Sipala,  V and Tosta e Melo,  I},
  year = {2023},
  month = Apr,
  pages = {1943–1964}
}

@article{Punturo2010,
  title = {The {Einstein Telescope}: a third-generation gravitational
    wave observatory},
  volume = {27},
  ISSN = {1361-6382},
  url = {http://dx.doi.org/10.1088/0264-9381/27/19/194002},
  DOI = {10.1088/0264-9381/27/19/194002},
  number = {19},
  journal = {Classical and Quantum Gravity},
  publisher = {IOP Publishing},
  author = {Punturo,  M and Abernathy,  M and Acernese,  F and Allen,  B and Andersson,  N and Arun,  K and Barone,  F and Barr,  B and Barsuglia,  M and Beker,  M and Beveridge,  N and Birindelli,  S and Bose,  S and Bosi,  L and Braccini,  S and Bradaschia,  C and Bulik,  T and Calloni,  E and Cella,  G and Mottin,  E Chassande and Chelkowski,  S and Chincarini,  A and Clark,  J and Coccia,  E and Colacino,  C and Colas,  J and Cumming,  A and Cunningham,  L and Cuoco,  E and Danilishin,  S and Danzmann,  K and De Luca,  G and De Salvo,  R and Dent,  T and De Rosa,  R and Di Fiore,  L and Di Virgilio,  A and Doets,  M and Fafone,  V and Falferi,  P and Flaminio,  R and Franc,  J and Frasconi,  F and Freise,  A and Fulda,  P and Gair,  J and Gemme,  G and Gennai,  A and Giazotto,  A and Glampedakis,  K and Granata,  M and Grote,  H and Guidi,  G and Hammond,  G and Hannam,  M and Harms,  J and Heinert,  D and Hendry,  M and Heng,  I and Hennes,  E and Hild,  S and Hough,  J and Husa,  S and Huttner,  S and Jones,  G and Khalili,  F and Kokeyama,  K and Kokkotas,  K and Krishnan,  B and Lorenzini,  M and L\"{u}ck,  H and Majorana,  E and Mandel,  I and Mandic,  V and Martin,  I and Michel,  C and Minenkov,  Y and Morgado,  N and Mosca,  S and Mours,  B and M\"{u}ller–Ebhardt,  H and Murray,  P and Nawrodt,  R and Nelson,  J and Oshaughnessy,  R and Ott,  C D and Palomba,  C and Paoli,  A and Parguez,  G and Pasqualetti,  A and Passaquieti,  R and Passuello,  D and Pinard,  L and Poggiani,  R and Popolizio,  P and Prato,  M and Puppo,  P and Rabeling,  D and Rapagnani,  P and Read,  J and Regimbau,  T and Rehbein,  H and Reid,  S and Rezzolla,  L and Ricci,  F and Richard,  F and Rocchi,  A and Rowan,  S and R\"{u}diger,  A and Sassolas,  B and Sathyaprakash,  B and Schnabel,  R and Schwarz,  C and Seidel,  P and Sintes,  A and Somiya,  K and Speirits,  F and Strain,  K and Strigin,  S and Sutton,  P and Tarabrin,  S and Th\"{u}ring,  A and van den Brand,  J and van Leewen,  C and van Veggel,  M and van den Broeck,  C and Vecchio,  A and Veitch,  J and Vetrano,  F and Vicere,  A and Vyatchanin,  S and Willke,  B and Woan,  G and Wolfango,  P and Yamamoto,  K},
  year = {2010},
  month = sep,
  pages = {194002}
}

@article{Saccorotti2023,
  title = {Array analysis of seismic noise at the {Sos Enattos} mine,
    the {Italian} candidate site for the {Einstein Telescope}},
  volume = {138},
  ISSN = {2190-5444},
  url = {http://dx.doi.org/10.1140/epjp/s13360-023-04395-2},
  DOI = {10.1140/epjp/s13360-023-04395-2},
  number = {9},
  journal = {The European Physical Journal Plus},
  publisher = {Springer Science and Business Media LLC},
  author = {Saccorotti,  Gilberto and Giunchi,  Carlo and D’Ambrosio,  Michele and Gaviano,  Sonja and Naticchioni,  Luca and D’Urso,  Domenico and Rozza,  Davide and Cardini,  Alessandro and Contu,  Andrea and Dordei,  Francesca and Cadeddu,  Matteo and Tuveri,  Matteo and Migoni,  Carlo and Punturo,  Michele and Allocca,  Annalisa and Calloni,  Enrico and Cardello,  Giovanni Luca and D’Onofrio,  Luca and Davari,  Nazanin and Dell’Aquila,  Daniele and De Rosa,  Rosario and Carpinelli,  Massimo and Di Fiore,  Luciano and di Giovanni,  Matteo and Errico,  Luciano and Fiori,  Irene and Tringali,  Maria Concetta and Harms,  Jan and Koley,  Soumen and Longo,  Vittorio and Majorana,  Ettore and Mangano,  Valentina and Olivieri,  Marco and Paoletti,  Federico and Pesenti,  Luca and Puppo,  Paola and Rapagnani,  Piero and Razzano,  Massimiliano and Ricci,  Fulvio and Sipala,  Valeria and Tosta e Melo,  Iara and Trozzo,  Lucia},
  year = {2023},
  month = sep
}

@article{Naticchioni2020,
  title = {Characterization of the {Sos Enattos} site for the
    {Einstein Telescope}},
  volume = {1468},
  ISSN = {1742-6596},
  url = {http://dx.doi.org/10.1088/1742-6596/1468/1/012242},
  DOI = {10.1088/1742-6596/1468/1/012242},
  number = {1},
  journal = {Journal of Physics: Conference Series},
  publisher = {IOP Publishing},
  author = {Naticchioni,  L and Boschi,  V and Calloni,  E and Capello,  M and Cardini,  A and Carpinelli,  M and Cuccuru,  S and D’Ambrosio,  M and de Rosa,  R and Di Giovanni,  M and d’Urso,  D and Fiori,  I and Gaviano,  S and Giunchi,  C and Majorana,  E and Migoni,  C and Oggiano,  G and Olivieri,  M and Paoletti,  F and Paratore,  M and Perciballi,  M and Piccinini,  D and Punturo,  M and Puppo,  P and Rapagnani,  P and Ricci,  F and Saccorotti,  G and Sipala,  V and Tringali,  M C},
  year = {2020},
  month = Feb,
  pages = {012242}
}

@article{LeGonidec2019,
  title = {Abrupt changes of hydrothermal activity in a lava dome detected by combined seismic and muon monitoring},
  volume = {9},
  ISSN = {2045-2322},
  url = {http://dx.doi.org/10.1038/s41598-019-39606-3},
  DOI = {10.1038/s41598-019-39606-3},
  number = {1},
  journal = {Scientific Reports},
  publisher = {Springer Science and Business Media LLC},
  author = {Le Gonidec,  Y. and Rosas-Carbajal,  M. and Bremond d’Ars,  J. de and Carlus,  B. and Ianigro,  J.-C. and Kergosien,  B. and Marteau,  J. and Gibert,  D.},
  year = {2019},
  month = Feb 
}

@article{Tramontini2024,
  title = {Defining the sensitivity of cosmic ray muons to groundwater storage changes},
  volume = {356},
  ISSN = {1778-7025},
  url = {http://dx.doi.org/10.5802/crgeos.277},
  DOI = {10.5802/crgeos.277},
  number = {G1},
  journal = {Comptes Rendus. Géoscience},
  publisher = {MathDoc/Centre Mersenne},
  author = {Tramontini,  Matías and Rosas-Carbajal,  Marina and Zyserman,  Fabio Iván and  Longuevergne,  Laurent and Nussbaum,  Christophe and Marteau,  Jacques},
  year = {2024},
  month = Nov,
  pages = {177–194}
}

@article{Tramontini2026,
  title = {Insights into the internal structure and mechanical behavior
    of {Copahue} volcano ({Argentina}–{Chile})},
  volume = {476},
  ISSN = {0377-0273},
  url = {http://dx.doi.org/10.1016/j.jvolgeores.2026.108644},
  DOI = {10.1016/j.jvolgeores.2026.108644},
  journal = {Journal of Volcanology and Geothermal Research},
  publisher = {Elsevier BV},
  author = {Tramontini,  Matías and Rosas-Carbajal,  Marina and Heap,  Michael J. and Besson,  Pascale and Marteau,  Jacques and Garcia,  Sebastian and Zyserman,  Fabio I.},
  year = {2026},
  month = Aug,
  pages = {108644}
}

@article{Bonechi2020,
  title = {Atmospheric muons as an imaging tool},
  volume = {5},
  ISSN = {2405-4283},
  url = {http://dx.doi.org/10.1016/j.revip.2020.100038},
  DOI = {10.1016/j.revip.2020.100038},
  journal = {Reviews in Physics},
  publisher = {Elsevier BV},
  author = {Bonechi,  Lorenzo and D’Alessandro,  Raffaello and Giammanco,  Andrea},
  year = {2020},
  month = nov,
  pages = {100038}
}

@misc{hamming1998digital,
  title={Digital Filters, {Mineola}},
  author={Hamming, RW},
  year={1998},
  publisher={New York: Dover Publications}
}

@article{Marteau2014,
  title = {Implementation of sub-nanosecond time-to-digital convertor in field-programmable gate array: applications to time-of-flight analysis in muon radiography},
  volume = {25},
  ISSN = {1361-6501},
  url = {http://dx.doi.org/10.1088/0957-0233/25/3/035101},
  DOI = {10.1088/0957-0233/25/3/035101},
  number = {3},
  journal = {Measurement Science and Technology},
  publisher = {IOP Publishing},
  author = {Marteau,  Jacques and d’Ars,  Jean de Bremond and Gibert,  Dominique and Jourde,  Kevin and Gardien,  Serge and Girerd,  Claude and Ianigro,  Jean-Christophe},
  year = {2014},
  month = feb,
  pages = {035101}
}

@article{Marteau2017,
  title = {{DIAPHANE}: muon tomography applied to volcanoes, civil
    engineering, archaelogy},
  volume = {12},
  ISSN = {1748-0221},
  url = {http://dx.doi.org/10.1088/1748-0221/12/02/C02008},
  DOI = {10.1088/1748-0221/12/02/c02008},
  number = {02},
  journal = {Journal of Instrumentation},
  publisher = {IOP Publishing},
  author = {Marteau,  J. and d’Ars,  J. de Bremond and Gibert,  D. and Jourde,  K. and Ianigro,  J.-C. and Carlus,  B.},
  year = {2017},
  month = feb,
  pages = {C02008–C02008}
}

@article{Marteau2012,
  title = {Muons tomography applied to geosciences and volcanology},
  volume = {695},
  ISSN = {0168-9002},
  url = {http://dx.doi.org/10.1016/j.nima.2011.11.061},
  DOI = {10.1016/j.nima.2011.11.061},
  journal = {Nuclear Instruments and Methods in Physics Research Section A: Accelerators,  Spectrometers,  Detectors and Associated Equipment},
  publisher = {Elsevier BV},
  author = {Marteau,  J. and Gibert,  D. and Lesparre,  N. and Nicollin,  F. and Noli,  P. and Giacoppo,  F.},
  year = {2012},
  month = dec,
  pages = {23–28}
}

@article{Bajou2023,
  title = {High-resolution structural imaging of volcanoes using improved muon tracking},
  volume = {235},
  ISSN = {1365-246X},
  url = {http://dx.doi.org/10.1093/gji/ggad269},
  DOI = {10.1093/gji/ggad269},
  number = {2},
  journal = {Geophysical Journal International},
  publisher = {Oxford University Press (OUP)},
  author = {Bajou,  R and Rosas-Carbajal,  M and Tonazzo,  A and Marteau,  J},
  year = {2023},
  month = jul,
  pages = {1138–1149}
}

@article{Lesparre2010,
  title = {Geophysical muon imaging: feasibility and limits},
  volume = {183},
  ISSN = {0956-540X},
  url = {http://dx.doi.org/10.1111/j.1365-246X.2010.04790.x},
  DOI = {10.1111/j.1365-246x.2010.04790.x},
  number = {3},
  journal = {Geophysical Journal International},
  publisher = {Oxford University Press (OUP)},
  author = {Lesparre,  N. and Gibert,  D. and Marteau,  J. and Déclais,  Y. and Carbone,  D. and Galichet,  E.},
  year = {2010},
  month = oct,
  pages = {1348–1361}
}

\end{document}